\documentclass[aps,prx,reprint,
               amsmath,amssymb,amsfonts,longbibliography,
               notitlepage]{revtex4-2}

\makeatletter
\g@addto@macro\bfseries{\boldmath}
\makeatother

\usepackage[varg]{txfonts}
\usepackage[T1]{fontenc}
\usepackage{color}
\usepackage{graphicx}
\graphicspath{{figures/}}

\usepackage{hyperref}
\definecolor{plotblue}{rgb}{0.1333,0.3686,0.6588}
\hypersetup{colorlinks=true, linkcolor=plotblue, citecolor=plotblue, urlcolor=plotblue}
\usepackage[capitalise]{cleveref}
\crefname{equation}{Eq.}{Eqs.}
\crefname{figure}{Fig.}{Figs.}
\crefname{section}{Sec.}{Secs.}
\crefname{appendix}{Appendix}{Appendices}
\crefname{subappendix}{Appendix}{Appendices}
\Crefname{equation}{Equation}{Equations}
\Crefname{figure}{Figure}{Figures}
\Crefname{section}{Section}{Sections}
\Crefname{appendix}{Appendix}{Appendices}
\Crefname{subappendix}{Appendix}{Appendices}
\newcommand{\subref}[2]{\cref{#1}\hyperref[#1]{#2}}

\renewcommand{\vec}[1]{\boldsymbol{#1}}
\newcommand{\vhat}[1]{\vec{\hat{#1}}}
\newcommand{\avg}[1]{\langle #1 \rangle}
\newcommand{\del}{\partial}
\newcommand{\J}{J_{0}}

\newcommand{\muval}{0.478}
\newcommand{\vbval}{1.642}

\begin{document}

\title{Seeding and Amplification of Chaos in Long-Range Interacting Spin Chains}
\author{Sareen Sabra}
\author{Jeffrey G. Rau}
\affiliation{ Department of Physics, University of Windsor, 401 Sunset Avenue,
    Windsor, Ontario, N9B 3P4, Canada }

\begin{abstract}
Truncating weak interactions is a natural simplification of a many-body
Hamiltonian, but the validity of this approximation need not be uniform in
distance and time. In a chaotic system, a weak long-range coupling can seed a
small disturbance far from its source, after which the local dynamics can
amplify that disturbance exponentially. We study this mechanism in an
infinite-temperature classical Heisenberg spin chain with a dominant
nearest-neighbor exchange and a weak interaction tail. Using a classical
decorrelator, we compare exponentially and power-law decaying tails over several
decades in interaction strength. For exponentially decaying interactions, the
long-distance front remains ballistic when the tail is sufficiently shallow,
with a velocity controlled by the competition between the spatial decay of the
seed and the local chaotic growth. For sufficiently rapid exponential decay, the
nearest-neighbor front, which is defined by an intrinsic butterfly  velocity
$v_b$, is recovered within the accessible space-time window. For power-law
interactions, the arrival time is logarithmic in distance for all exponents
studied, corresponding to an exponentially expanding spatial front; the
power-law exponent controls the coefficient of this logarithm. We interpret both
behaviors using a seeding-and-amplification picture in which long-range
couplings generate weak disturbances throughout the chain whose growth is
governed primarily by the local nearest-neighbor dynamics. In this picture, the
overall strength of the tail primarily controls when the long-range front
becomes visible, while its spatial decay controls the propagation law. Thus a
weak interaction tail can leave conventional spin-dynamical observables
essentially unchanged while qualitatively altering the long-distance propagation
of decorrelation.
\end{abstract}

\maketitle

\section{Introduction}
\label{sec:introduction}

Chaotic dynamics combines local exponential sensitivity to initial conditions
with the spatial propagation of that sensitivity through an extended system.
Characterizing chaos in a spatially extended system therefore requires both a
growth rate and a propagation velocity. This idea was developed in the
spatiotemporal-chaos literature through spatial and comoving Lyapunov
spectra~\cite{lepri1996chronotopic}, and the velocity at which correlations
propagate was identified with a zero of the convective Lyapunov
exponent~\cite{giacomelli2000convective}. The same structure reappears in the
recent many-body-chaos literature as the velocity-dependent Lyapunov exponent,
whose zero defines the butterfly velocity~\cite{khemani2018velocity}. The
classical Heisenberg chain provides a particularly simple setting in which these
two aspects can be separated: at infinite temperature, a localized perturbation
grows rapidly while its influence spreads ballistically through the chain with a
finite butterfly velocity~\cite{das2018light}.  

The assumption of strict locality is, however, an idealization. Long-range
couplings are generic in the platforms now used to study many-body dynamics,
from dipolar molecules and Rydberg arrays to trapped
ions~\cite{defenu2023longrange}. Even when the dominant interaction is between
nearest neighbors, weaker couplings at longer distances provide a direct channel
by which a perturbation influences remote degrees of freedom. The effect of such
a tail can be parametrically small in the local equations of motion while
remaining important at the leading edge of a disturbance, where the decorrelator
itself is exponentially small. In a power-law quantum spin chain, for example,
the long-distance commutator is controlled by short-time perturbation theory in
the direct coupling~\cite{colmenarez2020lieb}, and in large-local-Hilbert-space
models the operator tail combines exponential temporal growth with an algebraic
spatial prefactor~\cite{chen2019quantum}. This raises a practical question,
closely related to the error analysis of truncated power-law interactions in
digital quantum simulation~\cite{tran2019locality}: when is truncating an
interaction a good approximation to the local dynamics but a poor approximation
to the spatial propagation of chaos? 

Related questions have a substantial history. Earlier studies of coupled-map
lattices examined how nonlocal couplings modify the propagation of an initially
localized disturbance~\cite{paladin1994predictability,torcini1997disturbance}.
Torcini and Lepri showed that in chaotic coupled-map lattices with algebraically
decaying coupling, disturbances propagate exponentially fast in time, with a
spatial growth rate set by the ratio of the maximal Lyapunov exponent to the
interaction exponent~\cite{torcini1997disturbance}. Classical
Lieb--Robinson-type bounds for long-range Hamiltonian lattices likewise admit
logarithmic causal regions, with numerical confirmation in classical
chains~\cite{metivier2014spreading}, and exact results are available for
classical out-of-time-order correlators in low-temperature power-law
models~\cite{kiran2021spatiotemporal}. On the quantum side, Chen and Zhou
obtained a logarithmic light cone for every finite interaction exponent in the
large-local-Hilbert-space limit~\cite{chen2019quantum}, whereas for finite local
Hilbert space the operator front depends sensitively on the exponent, with
power-law, linear, and intermediate light cones all
appearing~\cite{zhou2020operator,colmenarez2020lieb}, and rigorous bounds
constrain these
behaviors~\cite{fossfeig2015nearly,tran2019locality,kuwahara2020strictly}.
Earlier work has therefore established that algebraically decaying couplings can
combine with local chaotic amplification to produce rapidly propagating
disturbances. Here our focus is different: starting from a fixed, strongly
chaotic nearest-neighbor Hamiltonian, we ask how a parametrically weak
interaction tail modifies the chaos front. This setup separates the overall
amplitude of the tail from its spatial profile and allows us to relate the
resulting front quantitatively to the growth and spreading of perturbations in
the nearest-neighbor chain.

The resulting physical picture is one of seeding and amplification. A
perturbation at the origin changes the exchange field at a distant site by an
amount proportional to the corresponding interaction. At short times this
generates a small local perturbation proportional to the long-range coupling.
Once generated, that perturbation is amplified and propagated primarily by the
same local chaotic dynamics that governs the nearest-neighbor chain. This is the
mechanism underlying the exponentially fast propagation of
\citet{torcini1997disturbance} and the algebraic-times-exponential operator tail
of \citet{chen2019quantum}. It is related in spirit to the secondary light cones
produced by scattering events in classical spin
dynamics~\cite{bilitewski2021classical,ruidas2026emergence}, although the seeds
studied here arise directly from the interaction tail rather than from
quasiparticle scattering. It also connects naturally to the description of chaos
fronts as propagating waves in many-body
systems~\cite{aleiner2016microscopic,xu2019locality} and to the general theory
of front selection by linearly amplified
perturbations~\cite{vansaarloos2003front}.

In this work we test this picture numerically for a classical Heisenberg chain
at infinite temperature. We consider exponentially and power-law decaying
exchange tails with amplitudes many orders of magnitude smaller than the
nearest-neighbor exchange, allowing their overall strength and spatial decay to
be varied independently. The parameters describing the growth and spreading of
perturbations are fixed entirely from the nearest-neighbor chain. The comparison
with the long-range simulations therefore introduces no additional fitted
parameters. We further validate this picture with simulations that include an
interaction at only a single additional distance, which show the seeding and its
subsequent amplification more directly. For an exponential tail, a continuum
envelope calculation predicts a ballistic long-range front when the decay is
sufficiently shallow and identifies a crossover to the nearest-neighbor
butterfly velocity when the tail becomes sufficiently steep. For a power-law
tail, the same argument gives a logarithmic arrival time so that the spatial
extent grows exponentially in time. The amplitude of the tail primarily shifts
the arrival time, while its spatial decay controls the predicted asymptotic
front.

The remainder of this paper is organized as follows.
\Cref{sec:model-and-methods} introduces the model, the decorrelator, and the
numerical methods. \Cref{sec:seeding-picture} develops the
seeding-and-amplification picture, which we then compare with simulations for
exponential (\cref{sec:exp-decay}) and power-law (\cref{sec:power-decay}) tails.
We discuss the implications and limitations in \cref{sec:discussion} and
conclude in \cref{sec:conclusion}. \Cref{sec:coupling-cutoff} justifies
truncating the interaction at $d=N-1$, \cref{sec:single-distance} tests the
seeding picture directly, and \cref{sec:error-minimization,sec:nn-analysis}
document the numerical checks and the nearest-neighbor calibration.

\section{Model and Methods}
\label{sec:model-and-methods}

\subsection{Model}
\label{sec:model}
We consider a one-dimensional classical ferromagnetic spin chain of length $N$.
Each spin is a three-dimensional unit vector $\vec{S}_x$, with
$|\vec{S}_x|^2=1$, and periodic boundary conditions imposed through
$\vec{S}_{x+N}=\vec{S}_x$. The Hamiltonian is
\begin{equation}
    H = -J_1 \sum_{x=1}^N \vec{S}_x \cdot \vec{S}_{x+1} - 
    \sum_{x=1}^N \sum_{d=2}^{N-1} J_d \vec{S}_x \cdot \vec{S}_{x+d}. 
    \label{eq:hamiltonian}
\end{equation}
Here $J_1 > 0$ is the dominant nearest-neighbor exchange and $J_d$ is a weak
tail. Since the displacement $d$ is defined modulo $N$, an unordered pair on the
ring receives contributions associated with both $d$ and $N-d$. For the system
sizes and decay parameters considered below, the latter is exponentially or
algebraically suppressed over the spatial region used to extract the front; see
\cref{sec:coupling-cutoff}.

We consider two classes of interaction tails. For exponentially decaying
interactions, we take $J_d$ to have the form
\begin{equation}
    J_d = 
    \begin{cases}
    J_1 & d=1, \\
    \J e^{-\alpha (d-2)} & d>1,
    \end{cases},
    \label{eq:exp-coupling}
\end{equation}
where $\alpha$ is the inverse decay length and $\J$ is the strength of the
largest further-neighbor coupling, $J_2=\J$. For power-law interactions,
\begin{equation}
    J_d = 
    \begin{cases}
    J_1 & d=1, \\
    \displaystyle \frac{\J}{(d-1)^\beta} & d>1,
    \end{cases},
    \label{eq:power-coupling}
\end{equation}
where $\beta$ is the power-law exponent and $J_2=\J$. We restrict attention to
$\beta>1$, for which the corresponding one-dimensional infinite-image sum is
convergent~\cite{defenu2023longrange}. 

The dynamics of the chain follow the equation of motion 
\begin{equation}
    \frac{d \vec{S}_x}{dt} = \{\vec{S}_x, H\} \equiv  \vec{S}_x \times \vec{B}_x ,
    \label{eq:eom}
\end{equation}
where $\vec{B}_x$ is the exchange field 
\begin{equation}
    \vec{B}_x \equiv -\frac{\del H}{\del \vec{S}_x} = 
    J_1\left(\vec{S}_{x+1} + \vec{S}_{x-1}\right) + 
    \sum_{d=2}^{N-1} J_d \left(\vec{S}_{x+d} + \vec{S}_{x-d}\right),
    \label{eq:exchange-field}
\end{equation}
following directly from \cref{eq:hamiltonian}. Note that the nearest-neighbor
term is not included in the long-range sum.

All results in this work are at infinite temperature ($T/J_1=\infty$). The
initial spins are therefore sampled independently and uniformly from the unit
sphere. To define the spreading of chaos, we follow the framework outlined by
\citet{das2018light} for nearest-neighbor spin chains. For each realization we
first initialize two identical copies, $\vec{S}_x$ and $\vec{S}'_x$. A small
perturbation is then introduced to the central spin of the second copy,
$\vec{S}'_{x_0}$, where $x_0 \equiv N/2$. Below, positions are measured relative
to $x_0$, so that the perturbed site is $x=0$. Explicitly, we perturb the spin
$\vec{S}'_{x_0}$ through a rotation by a small angle $\epsilon$ about a randomly
chosen axis, $\vhat{n}$. The two chains are then separately evolved in time
using the (deterministic) equation of motion [\cref{eq:eom}]. The small
difference at $x=x_0$ causes the two trajectories to diverge, with the degree of
divergence depending on both position and time.

To quantitatively characterize their divergence we use the decorrelator
introduced for the nearest-neighbor chain in Ref.~\cite{das2018light},
\begin{equation}
    D(x,t) \equiv 1- \avg{\vec{S}^{}_x(t) \cdot \vec{S}'_x(t)} = 
    \frac{1}{2}\avg{|\vec{S}_x(t)-\vec{S}'_x(t)|^2}.
    \label{eq:decorrelator}
\end{equation}
Small values $D(x,t)\ll1$ correspond to nearly identical trajectories; the spins
$\vec{S}_x$ and $\vec{S}'_x$ are highly correlated. Larger values $D(x,t) =
O(1)$ signal substantial decorrelation with
$\avg{\vec{S}^{}_x\cdot\vec{S}'_x}\approx 0$, indicating that the influence of
the initial perturbation has spread to the site $x$. This decorrelator is a
classical counterpart of the out-of-time-order correlator (OTOC) used to
characterize chaos in quantum mechanical
systems~\cite{das2018light,khemani2018velocity}.

To characterize the arrival of the perturbation at a given site, we define the
arrival time at threshold $\chi$ by
\begin{equation}
    \tau(x)=\min\{t:D(x,t)\geq\chi\}.
    \label{eq:arrival-time}
\end{equation}
Unless otherwise stated, we use $\chi=10^{-3}$. This lies well above the initial
decorrelator, $D(0,0)\simeq\epsilon^2/3$ [\cref{eq:initial-decorrelator}], while
remaining in the small-decorrelator regime in which the nearest-neighbor growth
form provides a useful description. The dependence on the perturbation amplitude
and the normalization of the nearest-neighbor growth form is discussed in
\cref{sec:nn-analysis}.

\subsection{Numerical Methods}
\label{sec:numerical-methods}
We proceed numerically, integrating \cref{eq:eom} with a fixed-step fourth-order
Runge--Kutta (RK4) method, which is convenient for advancing both copies on a
common time grid.\footnote{A common fixed time grid ensures that the two copies
are advanced using identical time steps. An adaptive method would instead
require the two copies to be integrated as a single combined system so that they
share the same accepted steps.} The calculations use multiprecision
floating-point arithmetic~\cite{boost}. The working precision was increased
until further increases produced no detectable change in the decorrelator or
extracted arrival times; for some parameter sets, up to 100 decimal digits were
required. The results shown below use $N=1024$, $\delta t=0.025$, and
$\epsilon=10^{-3}$. We set $J_1=1$, so that energies are measured in units of
$J_1$ and times in units of $J_1^{-1}$. Each parameter set is averaged over
$6000$ independently sampled initial conditions and perturbation axes, and any
statistical uncertainties are estimated by bootstrap resampling of complete
realizations.~\footnote{ The original simulation produced a symmetric front
extending from $x=0$ to $x=1024$, with the first peak emerging at the center of
the chain, $x=512$. We exploit the $x \rightarrow -x$ symmetry by ``folding''
the data across $x=0$ to increase the effective number of samples. }

We have verified that this time step is sufficiently small that the effects of
discretization error on our results are minimal.  We assess time-step
convergence at nonzero times by comparing $D(x,t)$ for several values of $\delta
t$. For details on the effect of the time step on the decorrelator, see
\cref{sec:error-minimization}, which also discusses numerical convergence and
the initialization checks.

We apply these methods to simulate spin chains with several different choices of
$J_d$ [\cref{eq:exp-coupling,eq:power-coupling}]. We study interaction scales
$\J/J_1=10^{-3}$, $10^{-5}$, and $10^{-7}$. For the exponential tail we consider
$\alpha=0.1,0.2,0.3,0.5,$ and $0.7$ (not shown), while for the power-law tail we
use $\beta=3,5,$ and $7$. We keep $\J$ small relative to $J_1$ to test how weak
long-range interactions affect the spread of chaos. The case $\beta=3$ shares
the $1/r^3$ spatial decay of dipole--dipole interactions, as realized, for
example, in dipolar spin exchange between lattice-confined polar
molecules~\cite{yan2013observation}\footnote{Note that the isotropic exchange
term in \cref{eq:hamiltonian} should not be identified with the full anisotropic
dipole--dipole interaction.}. The arrival time, $\tau(x)$, is used to map the
``front'' of the propagation of chaos, marking the boundary between regions of
correlation and decorrelation between the two copies of the system.

\section{Results}
\label{sec:results}

We first establish the physical mechanism underlying the behavior of the front
with long-range interactions and then apply it to exponential and power-law
tails. The central observation is that the tail need not dominate the local
dynamics in order to dominate the leading edge of the decorrelator. A very weak
interaction generates a correspondingly weak disturbance at a distant site, but
that disturbance is subsequently amplified by the chaotic dynamics.

\subsection{Conceptual Picture: Short-Time Seeding and Nearest-Neighbor Cones}
\label{sec:seeding-picture}
\begin{figure}[t!]
    \centering
    \includegraphics[width=\columnwidth]{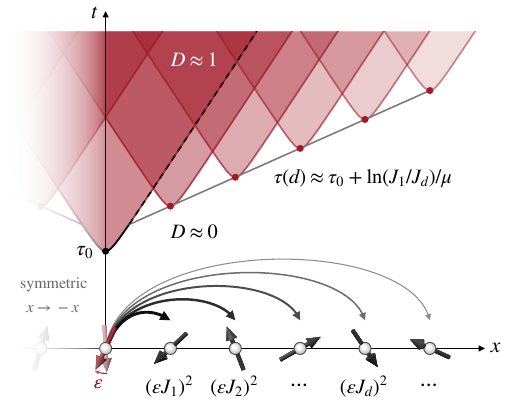}
\caption{Illustration of the seeding mechanism. The perturbation $\epsilon$ at
the origin seeds weak perturbations at distant sites through the long-range
couplings $J_d$. The decorrelator $D$ depends quadratically on these seed
perturbations, giving a factor $(\epsilon J_d)^2$. Each seed is subsequently
amplified by the local chaotic dynamics, producing a translated nearest-neighbor
cone with arrival time $\tau(d)$. The overlapping cones produce an extended
region of substantial decorrelation ($D\approx1$), separated from the
undecorrelated region ($D\approx0$) by the resulting long-range front.}
    \label{fig:seeded-cones}
\end{figure}

To model this phenomenon, we propose an analytic framework using the
nearest-neighbor front formula presented in Ref.~\cite{das2018light}. We treat
the front as a superposition of linear nearest-neighbor cones originating at
different spatial positions. Each of these cones is generated by a perturbation
whose strength and influence are governed by the weak long-range exchange
interactions (\cref{fig:seeded-cones}).

The scaling of the seed with $J_d$ follows directly from the linearized
equations of motion. Let $\delta\vec{S}_x \equiv \vec{S}'_x-\vec{S}_x$. Immediately
after the perturbation, $\delta\vec{S}_x(0)=0$ for $x\neq 0$. Linearizing
\cref{eq:eom} gives, schematically,
\begin{equation}
    \frac{d\,\delta\vec{S}_x}{dt}
    =\delta\vec{S}_x\times\vec{B}_x
    +\vec{S}_x\times \delta\vec{B}_x,
    \label{eq:linearized-eom}
\end{equation}
where $\delta\vec{B}_x=\sum_y K_{xy}\delta\vec{S}_y$ and $K_{xy}=J_d+J_{N-d}$,
with $d=(y-x)\bmod N$, is the exchange kernel implied by
\cref{eq:exchange-field}; away from the wrap-around region it reduces to
$J_{|x-y|}$. At a site coupled directly to the origin,
\begin{equation}
    \delta\vec{S}_x(t)
    =tK_{x0} \vec{S}_x(0)\times \delta\vec{S}_0(0)+O(t^2).
    \label{eq:short-time-seed}
\end{equation}
For independent infinite-temperature spins and a uniformly random rotation axis,
\cref{eq:short-time-seed} gives
\begin{equation}
    D(x,t)=\frac{2}{9}(\epsilon^2 K_{x0}^2)t^2+O(t^3,\epsilon^4),
    \label{eq:short-time-decorrelator}
\end{equation}
for $x \neq 0$. Thus the distant decorrelator is seeded at order
$\epsilon^2J_d^2$ based on this short-time expansion.

The immediate seeding of the decorrelator at sites coupled to the initial
perturbation by an exchange $J_d$ produces secondary cones that spread through
the dominant nearest-neighbor interaction.  For this growth we use the
phenomenological nearest-neighbor form motivated by Ref.~\cite{das2018light},
\begin{equation}
    D_1(x,t)
    =A\epsilon^2
    \exp\left[
        2\mu t\left(1-\frac{x^2}{(v_{b}t)^2}\right)
    \right],
    \label{eq:nn-growth-form}
\end{equation}
where $\mu$ is the local growth rate, $v_{b}$ is the nearest-neighbor butterfly
velocity, and $A$ is an empirical normalization. \Cref{eq:nn-growth-form} should
be viewed as a small-$D$ growth kernel rather than as an exact expression
throughout the nonlinear interior of the front~\cite{das2018light}.
\Cref{sec:nn-analysis} describes the calibration of this form against the
nearest-neighbor simulations. The parameters $A$ and $\mu$ are fitted to these
nearest-neighbor simulations. For the butterfly velocity we use the value
$v_b\approx\vbval$ reported by \citet{das2018light}, which is consistent with
our nearest-neighbor results. All three are held fixed throughout the analysis
of the long-range models, and no parameters are refitted to the long-range data.

Motivated by this short-time scaling [\cref{eq:short-time-decorrelator}], we
approximate the evolution of a seed generated at distance $d$ by a translated
nearest-neighbor growth kernel\footnote{ The short-time expansion
[\cref{eq:short-time-decorrelator}] fixes the dependence of the seed on $J_d$
but not its overall matching normalization; any $J_d$-independent matching
factor produces only an additive shift in the threshold arrival time and does
not affect the front slopes derived below.}, with an amplitude reduced by
$(J_d/J_1)^2$,
\begin{align}
    D_d(x,t)
    &\simeq A\epsilon^2\left(\frac{J_d}{J_1}\right)^2
    \exp\left[
        2\mu t\left(1-\frac{(x-d)^2}{v_{b}^2t^2}\right)
    \right] \nonumber\\ &=
    \left(\frac{J_d}{J_1}\right)^2 D_1(x-d,t)
    .
    \label{eq:seeded-cone}
\end{align}
The simplest approximation is that each of these seeded cones propagates
independently (at least until they overlap), giving the decorrelator
\begin{equation}
    D(x,t)
    \simeq D_{1}(x,t)+\sum_{d=2}^{N-1}D_d(x,t).
    \label{eq:seeded-sum}
\end{equation}
This is somewhat phenomenological: the true linearized perturbation vectors add
coherently, whereas \cref{eq:seeded-sum} adds their squared, ensemble-averaged
magnitudes. Cross terms and the continuous generation of seeds at later times
are neglected. The accuracy and usefulness of the approximation are therefore an
empirical question. At infinite temperature, the short-time seeds at distinct
sites have vanishing cross correlations because the initial spins are
independently and isotropically distributed. We therefore expect interference
between their subsequently amplified contributions to be suppressed by ensemble
averaging, although such terms need not vanish exactly once the cones overlap.
Repeated seeding introduces additional factors of the weak long-range exchange.
Each further transfer suppresses the decorrelator by another factor of the
corresponding $(J_d/J_1)^2$, delaying threshold arrival. We neglect these
repeated transfers in the leading approximation for the full interaction tails.
\Cref{sec:single-distance} provides a direct test using a single long-range
coupling, with $J_d\neq 0$ for only one distance $d\gg 1$. Since the seeded site
is itself coupled to a site a further distance $d$ away, this single coupling
generates a comb of translated nearest-neighbor cones at multiples of $d$.

A useful first approximation to the chaotic front is obtained by evaluating a
seeded cone at its center, $x=d$. Setting $D_d(d,t)=\chi$ gives
\begin{equation}
    \tau(d)
    \approx\frac{1}{2\mu}
    \ln\left(\frac{\chi J_1^2}{A\epsilon^2J_d^2}\right)
    = \tau_0 + \frac{1}{\mu}\ln\left(\frac{J_1}{J_d}\right)
    .
    \label{eq:seed-arrival-time}
\end{equation}
Here, $\tau_0\equiv \ln[\chi/(A\epsilon^2)]/(2\mu)$ is the approximate onset
time for the initial perturbation. We refer to this estimate as the arrival-time
approximation; it correctly captures the leading dependence on the spatial decay
of the seed, but considers only the center of each seeded cone. To account for
the spatial extent of the cones, we construct an approximate front from the
lower envelope of their individual threshold arrival times,
$\tau_{\rm env}(x)=\min_d\tau_d(x)$, where $\tau_d(x)$ is defined by
$D_d(x,\tau_d(x))=\chi$. This envelope is used in the figure comparisons below;
it is distinct from extracting the threshold of the full sum in
\cref{eq:seeded-sum}.

\subsection{Exponentially Decaying Interactions}
\label{sec:exp-decay}

\begin{figure}[htp]
    \centering
    \includegraphics[width=\columnwidth]{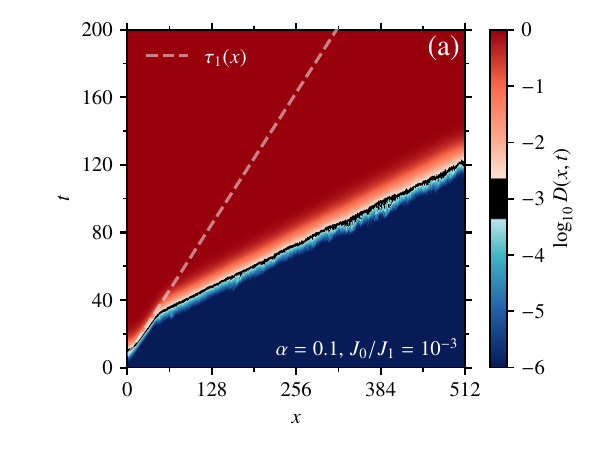}
    \includegraphics[width=\columnwidth]{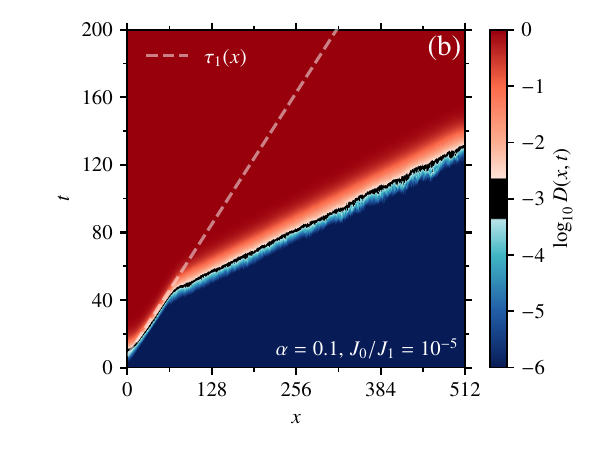}
    \includegraphics[width=\columnwidth]{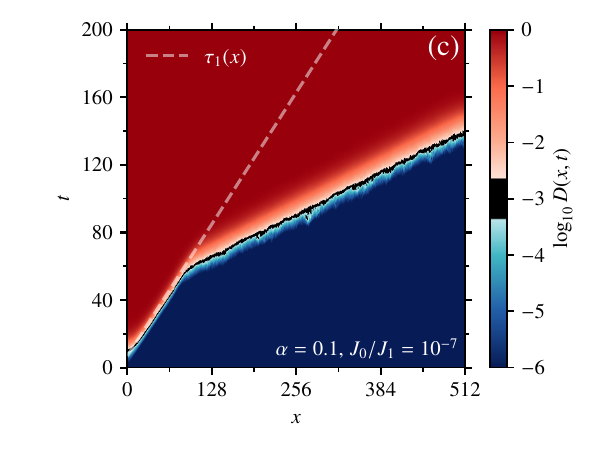}
    \caption{Simulated decorrelators, $D(x,t)$, for a perturbation of size
$\epsilon = 10^{-3}$ in a spin chain of length $N = 1024$ subjected to an
exponentially decaying $J_d$ with $\alpha=0.1$ and (a) $\J=10^{-3}$, (b)
$10^{-5}$, and (c) $10^{-7}$. The emphasized region (outlined in black)
indicates the boundary of correlation to decorrelation at $D(x,t) = \chi =
10^{-3}$, which defines the front of the propagation of chaos. At small times
the front is identical to the nearest-neighbor light cone, but at a crossover
time $t_c$ a linear tail emerges with a different slope. At fixed $\alpha$, the
crossover time shifts with the interaction amplitude $\J$.} 
    \label{fig:exp-colormaps}
\end{figure}

To compare this conceptual picture with our simulations, we begin with the case
of exponentially decaying interactions. Our results reveal that the light cone
for exponentially decaying interactions is composed of two distinct components:
the familiar nearest-neighbor cone governed by the butterfly velocity, and a
secondary cone defined by a linear front with an increased propagation velocity.
This secondary velocity depends on the interaction decay parameter $\alpha$ and
approaches $\mu/\alpha$ for small $\alpha$. Above a critical value $\alpha_c$,
derived below, the secondary front no longer emerges ahead of the
nearest-neighbor cone.

These results are presented in \cref{fig:exp-colormaps}, which shows the
decorrelator at $\alpha = 0.1$ for three coupling ratios: $J_0/J_1 = 10^{-3}$,
$10^{-5}$, and $10^{-7}$. For reference, the additional interaction strengths
for $J_0/J_1=10^{-3}$ are $J_2/J_1 = 10^{-3}$ at the second-neighbor distance,
while at distance $d=256$, $J_{256}/J_1$ drops to $9.3 \times 10^{-15}$. This
rapid decay highlights how quickly these interactions diminish relative to
$J_1$, even at large decay lengths. As \cref{fig:exp-colormaps} shows, despite
their small magnitude these long-range interactions have a significant effect on
the spread of chaos. At early times, the front follows the nearest-neighbor cone
with the butterfly velocity $v_b$ identified in Ref.~\cite{das2018light}. At
later times, a secondary front emerges and moves ahead of the original cone. 

Over the weak-tail range studied, changing $J_0$ primarily shifts the front to
later times, while its long-distance propagation velocity remains approximately
unchanged, as shown in \cref{fig:exp-colormaps}. A plausible explanation for
this behavior is that at early times the visible front is governed primarily by
the nearest-neighbor interactions because $J_0/J_1\ll1$. The long-range
interactions nevertheless seed weak perturbations at distant sites, which are
then amplified by the local chaotic dynamics, eventually producing the secondary
front. Fitting of the data suggests that the arrival-time shift at fixed $x$
scales as $\sim 2\ln(J_1/J_0)$, as one expects from this picture and
\cref{eq:nn-growth-form}. 

\begin{figure}[htp]
    \centering
    \begin{minipage}[t]{\columnwidth}
    \includegraphics[width=\textwidth]{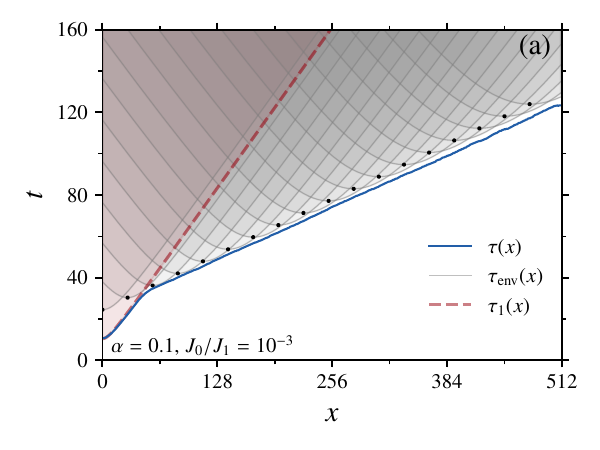}
    \end{minipage}
    \centering
    \begin{minipage}[t]{\columnwidth}
    \includegraphics[width=\textwidth]{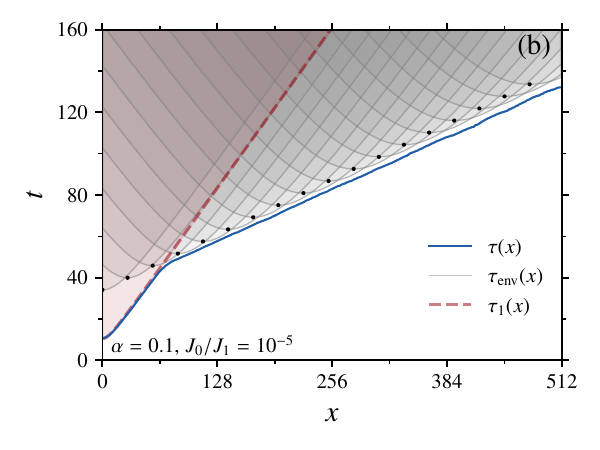}
    \end{minipage}
    \centering
    \begin{minipage}[t]{\columnwidth}
    \includegraphics[width=\textwidth]{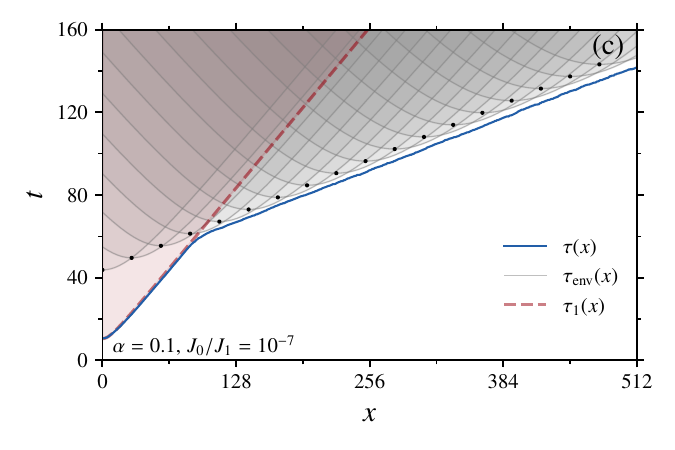}
    \end{minipage}
    \caption{Comparison of the seeded-cone approximation
    (\cref{sec:seeding-picture}) to the simulated data $\tau(x)$ for $\alpha =
    0.1$. The overlaid curves illustrate the front produced by the superposition
    of the nearest-neighbor cones from the seeded perturbations, $\tau_{\rm
    env}(x)$. The points indicate the minimum of each separately seeded cone,
    i.e., the earliest threshold arrival associated with that cone, given by the
    arrival-time approximation [\cref{eq:seed-arrival-time}]. Both estimates
    agree well with the simulated results, with the envelope providing the
    closer description of the front.}
    \label{fig:exp-envelopes}
\end{figure}

In \cref{fig:exp-envelopes}, we compare the seeded-cone approximation with the
simulated data for $\alpha=0.1$ across several values of $\J$. The superposed
nearest-neighbor cones closely reproduce the shape of the simulated front. In
particular, both the simulations and the approximation show a linear long-range
front whose slope is independent of $\J$, while the time at which this front
becomes visible shifts with the overall interaction strength. This indicates
that the propagation velocity is controlled primarily by the spatial decay
parameter $\alpha$, whereas $\J$ controls the temporal onset of the long-range
contribution.

We can understand this from the arrival time approximation. Following
\cref{eq:seed-arrival-time} we expect a secondary front to develop 
\begin{equation}
    \tau(d)
    \approx 
    \tau_0 + \frac{1}{\mu}\ln\left(\frac{J_1}{J_0}\right) + 
    \frac{\alpha}{\mu} (d-2).
    \label{eq:exp-seed-arrival-time}
\end{equation}
This is a linear front, as in the nearest-neighbor case, but with an
arrival-time shift of $\ln\left({J_1}/{J_0}\right)/\mu$. Using
$\mu\approx\muval$ from our nearest-neighbor calibration
(\cref{sec:nn-analysis}), the predicted arrival delay as a function of $\J$ is
$|\ln J_0|/\mu$ with coefficient $\mu^{-1}\approx\fpeval{round(1/\muval,2)}$
agreeing with the fitted value. This explains why changing the overall strength
of the tail can produce a large temporal delay without changing the functional
form of the long-distance front within the accessible window. Within this
arrival-time approximation, this new front spreads with the modified velocity
$\mu/\alpha$. 

Part of this new front is hidden by the nearest-neighbor cone. Since the initial
nearest-neighbor front follows $t\simeq \tau_0/2+d/v_b$ at large $d$, the two
fronts cross at
\begin{equation}
t_c =
\frac{
\left(\mu-\alpha v_b/2\right)\tau_0
+\ln\left(J_1/J_0\right)-2\alpha
}{
\mu-\alpha v_b
}.
\label{eq:exp-crossing-time}
\end{equation}
When $J_0\ll J_1$, the logarithm dominates, giving
\begin{equation}
t_c \approx
\frac{1}{\mu-\alpha v_b}
\ln\left(\frac{J_1}{J_0}\right).
\label{eq:exp-crossing-time-approx}
\end{equation}
With $v_b\approx\vbval$ from Ref.~\cite{das2018light}, this gives a coefficient
of approximately $\fpeval{round(1/(\muval-0.1*\vbval),2)}$ for $\alpha=0.1$. The
apparent divergence at $\alpha=\mu/v_b$ is an artifact of the arrival-time
approximation, which neglects overlap between the seeded cones.

This overlap can be treated more systematically. Replacing the sum in
\cref{eq:seeded-sum} by a continuum integral over the seed position $x'$ and
retaining only the exponential dependence gives
\begin{equation}
    D(x,t)\propto
    \int_0^\infty dx'\,
    \exp\left[
        -2\alpha x'+2\mu t\left( 1-\frac{(x-x')^2}{v_{b}^2t^2}\right)
    \right].
    \label{eq:exp-continuum}
\end{equation}
For large $x$ and $t$, the dominant contribution comes from the saddle point
\begin{equation}
    x_*=x-\frac{\alpha v_{b}^2t}{2\mu}.
    \label{eq:exp-saddle-point}
\end{equation}
When $x_*>0$, the leading exponent at fixed ray $x=vt$ is
\begin{equation}
    \frac{1}{t}\ln D
    =2\mu-2\alpha v+\frac{\alpha^2v_{b}^2}{2\mu}+O(t^{-1}).
    \label{eq:exp-ray-growth-rate}
\end{equation}
The threshold front is therefore ballistic with velocity
\begin{equation}
    v(\alpha)
    =\frac{\mu}{\alpha}
    +\frac{\alpha v_{b}^2}{4\mu},
    \label{eq:exp-front-velocity}
\end{equation}
Within the nearest-neighbor growth form of \cref{eq:nn-growth-form}, the
crossover occurs at $\alpha_c=2\mu/v_b$, where $v(\alpha_c)=v_b$. Using
$\mu\approx\muval$ (\cref{sec:nn-analysis}) and
$v_b\approx\vbval$~\cite{das2018light}, this gives $\alpha_c \approx
\fpeval{round(2*\muval/\vbval,3)}$, exactly twice the value $\mu/v_b$ predicted
from the first term alone. When $\alpha > \alpha_c$ the saddle point lies
outside the allowed integration region and the local nearest-neighbor front
controls the leading edge within this approximation. Note that the arrival-time
approximation, $v(\alpha)\simeq\mu/\alpha$, is recovered when $\alpha$ is small;
the second term in \cref{eq:exp-front-velocity} explains why this approximation
becomes less accurate as the exponential tail is made steeper.

\begin{figure}[htbp]
    \centering
    \includegraphics[width=\columnwidth]{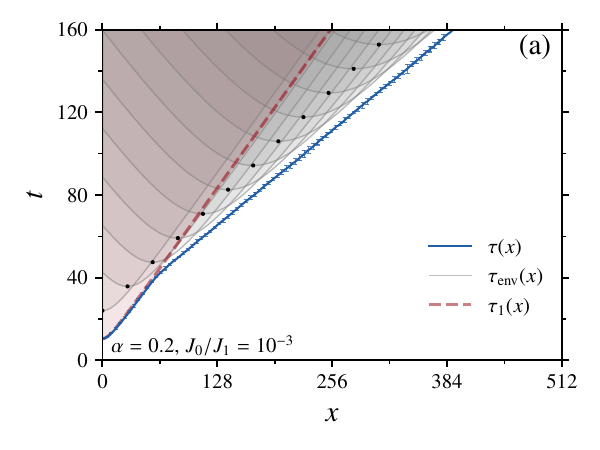}
    \includegraphics[width=\columnwidth]{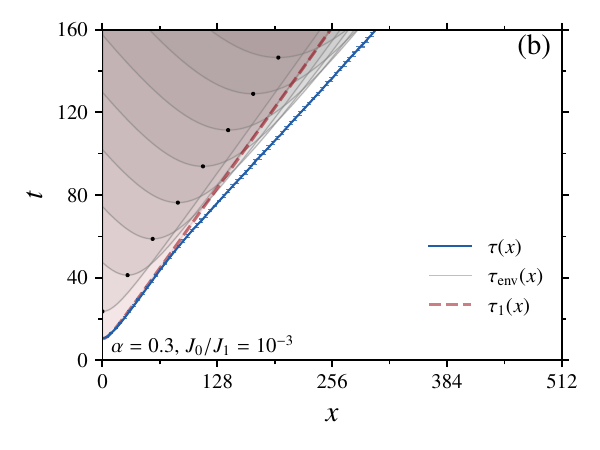}
    \includegraphics[width=\columnwidth]{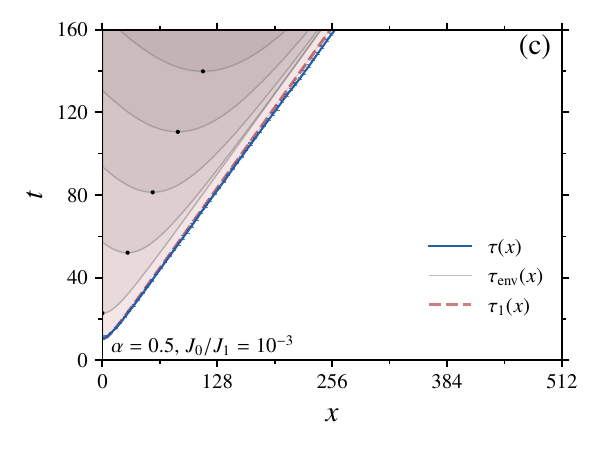}
    \caption{Comparison between the seeded-cone approximation
$\tau_{\rm env}(x)$ and the simulated results $\tau(x)$ for (a) $\alpha=0.2$,
(b) $\alpha=0.3$, and (c) $\alpha=0.5$ at fixed $\J=10^{-3}$. As $\alpha$
increases, the long-range front approaches the nearest-neighbor cone, while the
seeded-cone envelope agrees more closely with the simulated data. At $\alpha =
0.5$, the simulated front remains close to the nearest-neighbor front over the
accessible distances, although this finite-window agreement does not by itself
locate the crossover. The cone minima, indicated by the dots, no longer trace
the simulated data, highlighting the decreasing effectiveness of the
arrival-time approximation at larger $\alpha$.}
    \label{fig:exp-alpha-comparison}
\end{figure}

To test this, we consider a range of $\alpha = 0.2$, $0.3$, and $0.5$ with $J_0
= 10^{-3}$ and compare the numerically obtained front, a direct envelope of
nearest-neighbor cones [\cref{eq:seeded-sum}], and the prediction of
\cref{eq:exp-front-velocity}, as shown in \cref{fig:exp-alpha-comparison}. We
see strong agreement between the simulated data and the superposed seeded
nearest-neighbor cones. The envelope of the overlapping seeded cones traces the
smooth front observed in the simulations, producing a consistent slope.
Moreover, the data support the approximation given by
\cref{eq:exp-front-velocity}, as the slopes at larger $\alpha$ values are
noticeably steeper than in the $\alpha = 0.1$ case. As previously discussed,
this steepening arises from the reduced influence of distant-neighbor spins as
$\alpha$ increases, corresponding to a shorter decay length in the interaction
profile. At $\alpha = 0.5$, the simulated front remains close to the
nearest-neighbor front over the accessible distances. This finite-window
agreement does not by itself locate the crossover. For $\alpha=0.7$, which lies
above the predicted crossover $\alpha_c\simeq\fpeval{round(2*\muval/\vbval,3)}$,
the simulated front is indistinguishable from the nearest-neighbor front over
the accessible space-time window (not shown).

\subsection{Power-Law Decaying Interactions}
\label{sec:power-decay}
We now extend the framework developed for exponentially decaying interactions to
the case of power-law decaying interactions, defined in
\cref{eq:power-coupling}. As with the exponential case, we begin by asking
whether these interactions significantly influence the front, starting with
small values of the exponent $\beta$. Since power-law interactions decay more
slowly with distance than exponential interactions, we anticipate qualitative
differences in the emergent tail, even if the structural and conceptual features
remain similar. 

We begin with the prediction of the seeding picture. Using $J_d=J_0/(d-1)^\beta$
in \cref{eq:seed-arrival-time} gives
\begin{equation}
    \tau(d)
    \approx
    \tau_0 + \frac{1}{\mu}\ln\left(\frac{J_1}{J_0}\right)+\frac{\beta}{\mu}\ln(d-1).
    \label{eq:power-seed-arrival-time}
\end{equation}
Thus the seeded-cone approximation predicts a logarithmic dependence on $d$ in
the arrival time,
\begin{equation}
    \tau(d)\sim {\rm const.}+\frac{\beta}{\mu}\ln d.
    \label{eq:power-log-arrival}
\end{equation}
\Cref{eq:power-log-arrival} describes an exponentially expanding spatial front,
$d \sim e^{\mu t/\beta}$. The exponent $\beta$ does not change the logarithmic
functional form, but it does change its coefficient and hence the exponential
spatial growth rate.

The arrival-time approximation is better justified asymptotically for a
power-law tail than for an exponential tail. Optimizing the exponent of a
translated seeded cone (as in \cref{sec:exp-decay}) gives
\begin{equation}
    x-x_*\simeq\frac{\beta v_b^2 t}{2\mu x_*}.
    \label{eq:power-saddle-point}
\end{equation}
The physical solution for the seed position $x_*$ that dominates the
decorrelator is
\begin{equation}
    x_*=\frac{1}{2}\left(
    x+\sqrt{x^2-\frac{2\beta v_b^2t}{\mu}}
    \right).
    \label{eq:power-dominant-seed}
\end{equation}
Using \cref{eq:power-saddle-point}, the exponent evaluated at the saddle can be
written as
\begin{equation}
    \ln D(x,t)
    =
    2\mu t-2\beta\ln x_*
    -\beta\left(\frac{x-x_*}{x_*}\right)
    +\mathrm{const.}
    \label{eq:power-saddle-exponent}
\end{equation}
For the physical branch, $x/2\leq x_*\leq x$, so at fixed threshold this implies
$
    2\mu t \sim 2\beta\ln x
$
and therefore
\begin{equation}
    \tau(x)\sim \frac{\beta}{\mu}\ln x.
    \label{eq:power-saddle-arrival-time}
\end{equation}
up to sub-logarithmic corrections. Consequently $t \ll x^2$ at large $x$, and
the solution above gives $x_* \rightarrow x$. Thus the dominant seed approaches
the observation point itself, and the center-arrival approximation becomes
asymptotically exact at leading logarithmic order.

\begin{figure}[htp]
    \centering
    \includegraphics[width=\columnwidth]{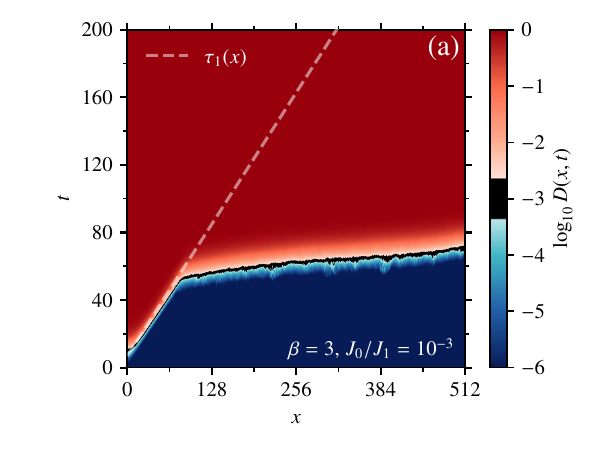}
    \includegraphics[width = \columnwidth]{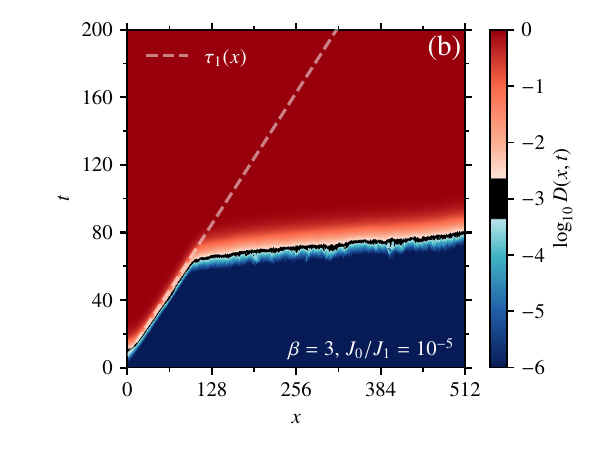}
    \includegraphics[width = \columnwidth]{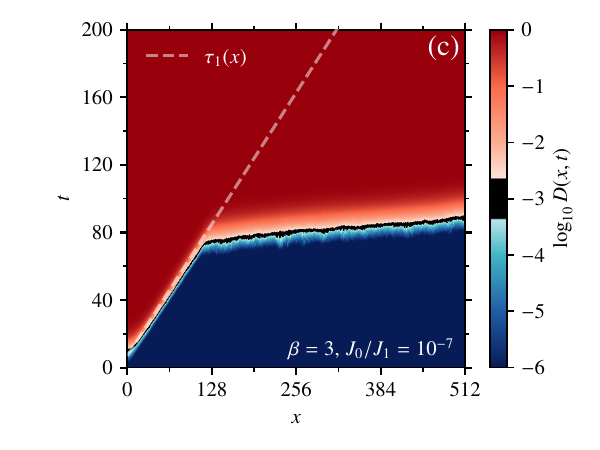}
    \caption{Simulated decorrelators, $D(x,t)$, for a perturbation of size
$\epsilon = 10^{-3}$ in a spin chain of length $N = 1024$ subjected to a
power-law decaying $J_d$ with $\beta=3$ and (a) $\J=10^{-3}$, (b) $10^{-5}$, and
(c) $10^{-7}$. The emphasized region (outlined in black) indicates the boundary
of correlation to decorrelation at $D(x,t) = \chi = 10^{-3}$, which defines the
front of the propagation of chaos. At small times the front is identical to the
nearest-neighbor light cone $\tau_1(x)$, but at a crossover time $t_c$ a
logarithmic tail emerges. At fixed $\beta$, the crossover time shifts with the
interaction amplitude $\J$.} 
    \label{fig:power-colormaps}
\end{figure}

\Cref{fig:power-colormaps} presents the simulated decorrelator for $\beta=3$ at
several values of $J_0/J_1$. As in the exponentially decaying case, the
long-range contribution does not immediately determine the visible front. At
early times the propagation is dominated by the nearest-neighbor dynamics, while
the weak power-law couplings seed perturbations at distant sites that
subsequently grow under the local chaotic dynamics. The long-range front becomes
visible only once these seeded perturbations reach the chosen decorrelator
threshold. Varying $J_0$ primarily shifts the arrival-time profile by
$\mu^{-1}\ln(J_1/J_0)$, analogous to the exponential case, without changing its
leading logarithmic dependence on distance.

\begin{figure}[htp]
    \centering
    \includegraphics[width=\columnwidth]{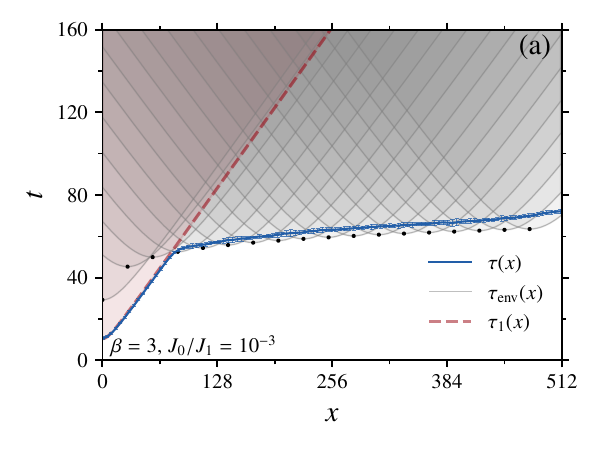}
    \centering
    \includegraphics[width=\columnwidth]{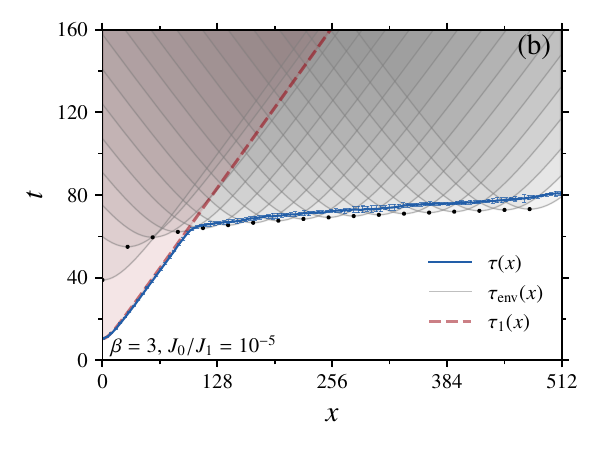}
    \centering
    \includegraphics[width=\columnwidth]{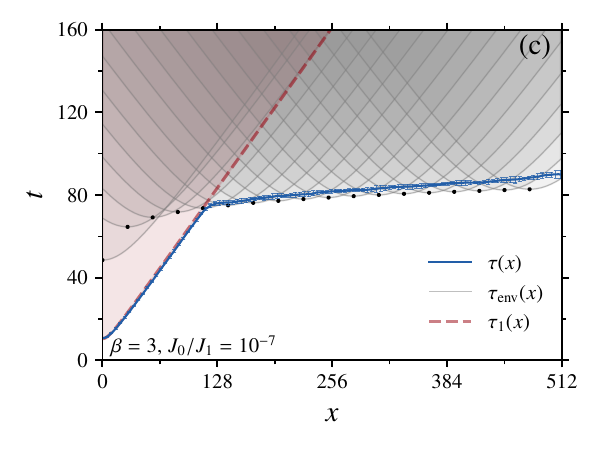}
    \caption{Power-law tails compared with the seeded-cone approximation, using
    exponent $\beta = 3$. The overlaid cones represent the front generated by
    superposing the nearest-neighbor cones from the seeded perturbations,
    $\tau_{\rm env}$. The dots indicate the minima of the individual seeded
    cones, the earliest threshold arrivals associated with each cone calculated
    using the arrival-time approximation [\cref{eq:seed-arrival-time}]. The
    approximation agrees closely with the simulated data, although some
    deviations emerge as $d \rightarrow N/2$, as observed in the exponentially
    decaying case.}
    \label{fig:power-envelopes}
\end{figure}

\begin{figure}[htp]
    \centering 
    \includegraphics[width=\columnwidth]{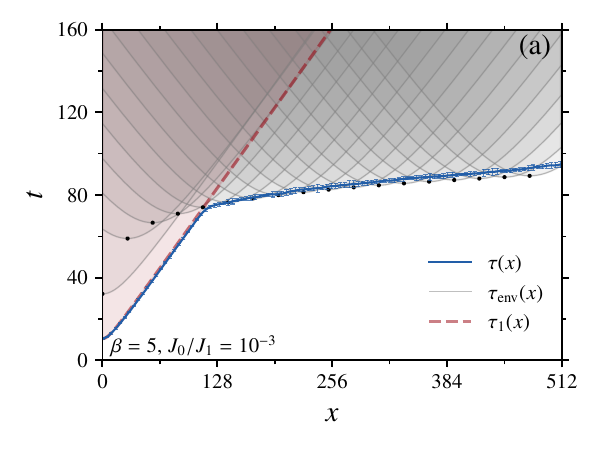}
    \centering   
    \includegraphics[width=\columnwidth]{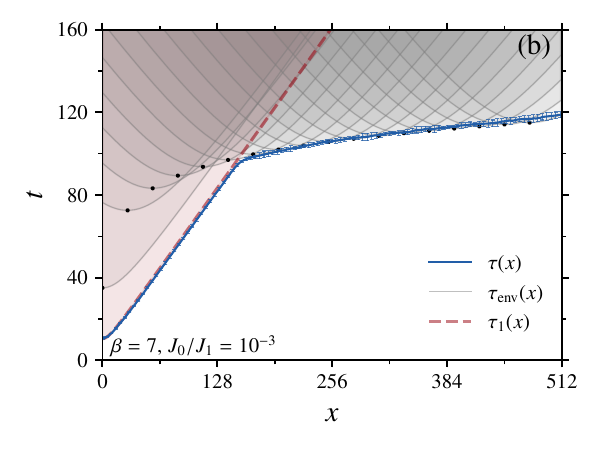}
    \caption{Power-law tails compared with the seeded-cone approximation for (a)
$\beta=5$ and (b) $\beta=7$, at fixed $\J=10^{-3}$. The minima stay closely
aligned with the boundary of both the simulated and seeded front. This suggests
that the arrival-time approximation continues to describe the front well for the
more rapidly decaying power-law interactions.}
    \label{fig:power-beta-comparison}
\end{figure}

\Cref{fig:power-envelopes} compares the seeded-cone approximation with the
$\beta=3$ simulations. The agreement supports the interpretation that the
long-range front is generated by weak spatial seeds followed by local chaotic
amplification. Finally, we compare faster power-law decays in
\cref{fig:power-beta-comparison}. The arrival-time profile remains nonballistic
and approximately logarithmic for $\beta=5$ and $7$. Increasing $\beta$ steepens
the arrival-time curve through the factor $\beta/\mu$ in
\cref{eq:power-seed-arrival-time}, while leaving its leading logarithmic
dependence on distance unchanged.

\section{Discussion}
\label{sec:discussion}
The simulations and the seeded-cone approximation point to a simple distinction
between conventional spin-dynamical observables and the long-distance
decorrelation front. When $\J/J_1\ll1$, the long-range tail produces only a
small correction to the local exchange field, so conventional spin-dynamical
observables can remain essentially unchanged from those of the nearest-neighbor
chain. The leading edge of the decorrelator is more sensitive. There, the
reference signal is itself exponentially small, so a directly seeded
perturbation at large distance can eventually dominate even when its coupling is
small. In this sense, truncating the interaction can be an excellent
approximation over a finite observation window while giving the wrong
long-distance propagation law.

The exponential and power-law cases differ because the spatial suppression of
the seed competes differently with temporal chaotic amplification. For an
exponential tail, both effects enter the exponent. The competition therefore
selects a finite ballistic velocity, \cref{eq:exp-front-velocity}, and produces
a crossover to the nearest-neighbor front when the tail is sufficiently steep.
For a power-law tail, the seed contributes only a logarithm to the arrival time.
The resulting relation $\tau(x)\propto\ln x$ corresponds to exponentially fast
growth of the front position. This general mechanism is consistent with earlier
observations of exponentially fast disturbance propagation in chaotic systems
with power-law couplings~\cite{torcini1997disturbance}, with the
algebraic-times-exponential operator tail of large-local-Hilbert-space quantum
models~\cite{chen2019quantum}, and with logarithmic causal regions in classical
long-range bounds~\cite{metivier2014spreading}. Our contribution is to connect
this behavior quantitatively to the independently characterized local growth
kernel in a weak-tail deformation of the classical Heisenberg chain, and to
separate the roles of tail amplitude and tail shape over several decades in
$\J$.

The interaction amplitude and spatial profile play distinct roles in this
picture. At fixed decay profile, decreasing $\J$ reduces the initial seeds and
primarily introduces an approximately logarithmic delay in threshold arrival,
without changing the leading propagation law. By contrast, changes in $\alpha$
or $\beta$  determine how the long-range front propagates. For exponential
interactions, $\alpha$ controls the velocity of the secondary ballistic front
[\cref{eq:exp-front-velocity}] and, more importantly, whether it can overcome
the nearest-neighbor front. For a power-law tail, the exponent controls the
coefficient $\beta/\mu$ of the logarithmic arrival time without changing its
functional form. The seeding argument predicts a logarithmic arrival time for
all $\beta>1$, and the simulations confirm this for $\beta=3$, $5$, and $7$.

Several limitations should be kept explicit. First, \cref{eq:seeded-sum} is not
a derivation of the full decorrelator. It neglects interference between
perturbation vectors, the continuous regeneration of seeds, and modifications of
the local growth kernel caused by the weak tail itself. The comparison with the
simulations suggests that these effects are subleading in the parameter regime
considered (as discussed in \cref{sec:seeding-picture}), but the approximation
should not be expected to describe the saturated interior of the front. Second,
all conclusions are drawn at finite $N$, finite threshold, and finite
observation time. For sufficiently small $\J$, the crossover can be pushed
beyond the useful numerical window even if the asymptotic tail would eventually
dominate. Conversely, at separations comparable to the circumference of the
ring, the distinction between direct and wrap-around propagation becomes
ambiguous (see \cref{sec:coupling-cutoff}).

There are several natural extensions. At finite temperature, the local growth
rate and propagation kernel become temperature
dependent~\cite{bilitewski2018temperature}, and in ordered regimes spin-wave
scattering can itself seed secondary
cones~\cite{bilitewski2021classical,ruidas2026emergence}. In higher dimensions
or with anisotropic exchange, the seeding profile becomes direction dependent
and should lead to anisotropic front selection.  Finally, comparison with
finite-local-Hilbert-space quantum systems requires care: quantum long-range
bounds and operator fronts depend sensitively on the interaction exponent and on
the diagnostic used~\cite{chen2019quantum,zhou2020operator,colmenarez2020lieb},
and for sufficiently summable power-law interactions the scrambling time of
thermodynamically stable quantum systems is bounded below by a power of the
system size~\cite{kuwahara2021absence}, in contrast to the exponentially
expanding classical front found here. Nonetheless, the framework we have
outlined could offer insights into the behavior of such quantum systems, through
this conceptual picture.

\section{Conclusion}
\label{sec:conclusion}

We have studied how weak interactions beyond nearest neighbors modify the
propagation of a local perturbation in an infinite-temperature classical spin
chain. The results support a seeding-and-amplification mechanism: the
interaction tail generates small disturbances at distant sites, and the local
chaotic dynamics subsequently amplifies and spreads those disturbances.
Exponential tails can produce a second ballistic front whose velocity is set by
the competition between spatial decay and temporal growth, while sufficiently
steep tails recover the nearest-neighbor butterfly front within the observed
window. Power-law tails produce logarithmic threshold arrival times,
corresponding to exponentially expanding spatial fronts, with a rate controlled
by the power-law exponent. The broader implication is not that weak long-range
interactions must always be retained, but that the validity of truncating them
depends on the observable, distance, and time scale of interest. A weak
interaction tail can leave conventional spin-dynamical observables essentially
unchanged while qualitatively altering the long-distance propagation of
decorrelation.

\begin{acknowledgments}
We thank S. Bhattacharjee for helpful comments. We acknowledge the support of
the Natural Sciences and Engineering Research Council of Canada (NSERC) (Funding
Reference No. RGPIN-2020-04970).
\end{acknowledgments}

\appendix
\crefalias{section}{appendix}  
\crefalias{subsection}{subappendix}
\section{Effect of Cutoff on Interactions}
\label{sec:coupling-cutoff}
In this appendix, we justify the long-range cutoff at $d=N-1$ and we show that
including periodic images of the interactions does not lead to a changed form of
the secondary tail. We stress that the two images at $d$ and $N-d$ are always
included.

Due to the chain's periodic boundary conditions, each position on the chain may
contribute an interaction to the local field $\vec{B}_x$ several times. The
corresponding effective coupling can be expressed
\begin{equation}
    \tilde{J}_d \equiv J_d + J_{d+N} + J_{d+2N} + \cdots = \sum_{m = 0}^\infty J_{d + mN}.
    \label{eq:periodic-coupling}
\end{equation}
We will show that these periodic copies have little effect on the spreading of
chaos for the cases of interest.

For exponentially decaying interactions we substitute \cref{eq:exp-coupling}
into \cref{eq:periodic-coupling}, yielding
\begin{equation}
    \tilde{J}_d = \sum_{m = 0}^\infty \J e^{-\alpha(d-2+mN)}
    = \J e^{-\alpha(d-2)}\left(\frac{1}{1 - e^{-\alpha N}}\right).
    \label{eq:periodic-exp-coupling}
\end{equation}
The form of the front depends on $\ln \tilde{J}_d$ and thus one has to consider
\begin{equation}
    \ln \tilde{J}_d = \ln J_d - \ln\left(1-e^{-\alpha N}\right).
    \label{eq:periodic-exp-log}
\end{equation}
For large system sizes, $\alpha N \gg 1$, so $\ln\left(1-e^{-\alpha N}\right)
\approx -e^{-\alpha N}$ and
\begin{equation}
    \ln \tilde{J}_d = \ln J_d + O(e^{-\alpha N}).
    \label{eq:periodic-exp-log-correction}
\end{equation}
These two forms thus agree up to exponentially small corrections; even for the
smallest decay rate studied, $\alpha=0.1$, $e^{-\alpha N}\approx
3\times10^{-45}$.

For power-law decay we substitute \cref{eq:power-coupling} into
\cref{eq:periodic-coupling}, yielding
\begin{equation}
    \tilde{J}_d = \sum_{m = 0}^\infty \frac{\J}{(d-1+mN)^\beta}.
    \label{eq:periodic-power-sum}
\end{equation}
For $\beta>1$ this sum converges. Separating the $m=0$ term gives
\begin{equation}
    \tilde{J}_d
    = J_d\left[
    1+\sum_{m=1}^{\infty}
    \left(\frac{d-1}{d-1+mN}\right)^\beta
    \right].
    \label{eq:periodic-power-coupling}
\end{equation}
For all $d$, the correction is bounded by
\begin{equation}
    \sum_{m=1}^{\infty}
    \left(\frac{d-1}{d-1+mN}\right)^\beta
    \leq
    \left(\frac{d-1}{N}\right)^\beta
    \sum_{m=1}^{\infty}\frac{1}{m^\beta}
    =
    \zeta(\beta)\left(\frac{d-1}{N}\right)^\beta ,
    \label{eq:periodic-power-bound}
\end{equation}
which is small for $d\ll N$. Hence
\begin{equation}
    \ln \tilde{J}_d
    =
    \ln J_d
    +O\left[\left(\frac{d}{N}\right)^\beta\right],
    \label{eq:periodic-power-log-correction}
\end{equation}
giving only algebraically small corrections in $d/N$.

Beyond this estimate, we evaluate the periodic-image correction numerically for
the parameters used in the simulations. For $N=1024$ and $2\leq d\leq N/2$, the
correction is largest at $d=N/2$, where for $\beta=3$, $5$, and $7$ it shifts
$\ln J_d$ by at most $0.0503$, $0.00448$, and $4.67\times10^{-4}$, respectively.
Periodic images thus change $J_d$ by at most about five percent for the slowest
decay considered, which by \cref{eq:seed-arrival-time} moves the predicted
arrival time by at most $0.0503/\mu\approx0.1$; for the larger values of $\beta$
the correction is negligible.

Therefore, for both exponential and power-law interactions, truncating the
interaction at $d=N-1$ preserves the spatial dependence of the coupling relevant
to the secondary front over the region used in our analysis. The omitted
periodic copies produce exponentially small corrections in the exponential case
and algebraically small finite-size corrections in the power-law case.

\section{Single-Distance Interaction}
\label{sec:single-distance}
\begin{figure}[htp]
    \centering
    \includegraphics[width=\columnwidth]{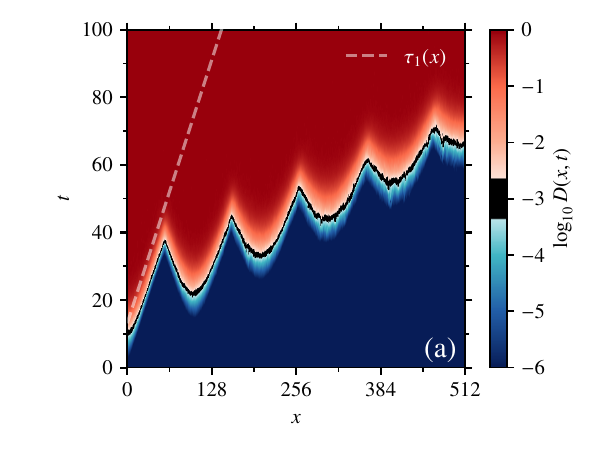}
    \includegraphics[width=\columnwidth]{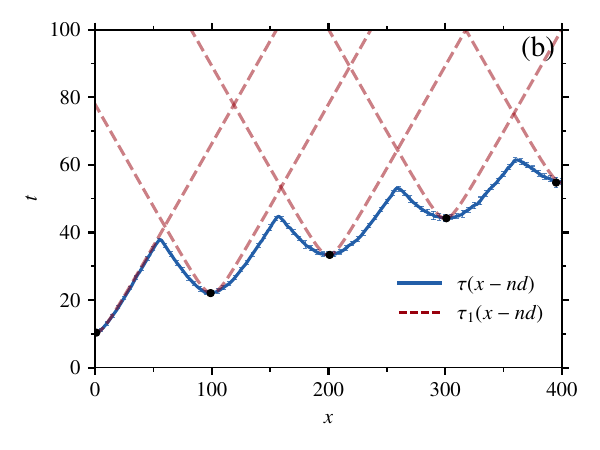}
    \caption{Decorrelator for a chain with a single long-range coupling,
    $J_d\neq 0$ only at $d=100$. (a) $\log_{10}D(x,t)$, with the
    nearest-neighbor boundary $\tau_1(x)$ shown dashed. (b) The recentered
    numerical fronts $\tau(x-nd)$ compared with the nearest-neighbor
    arrival-time profiles $\tau_1(x-nd)$ (dashed), for $nd=100$, $200$, $300$,
    and $400$. Repeated seeding through the same coupling produces a comb of
    cones at multiples of $d$; the faint lines have the nearest-neighbor slope
    and outline the resulting peaks.}
    \label{fig:single-distance}
\end{figure}

To isolate the seeding mechanism, we retain only the nearest-neighbor coupling
and a single long-range coupling $J_{100}=10^{-3}$ at $d=100$, setting
$J_{d'}=0$ for all other $d'>1$. The perturbation first seeds a cone at $x=d$;
that seeded site is in turn coupled to $x=2d$, and so on, producing a comb of
nearest-neighbor cones at $x=nd$ (\cref{fig:single-distance}). Since each
additional seeding step costs a further factor of $(J_d/J_1)^2$ in the
decorrelator, \cref{eq:seed-arrival-time} predicts that successive cone minima
are separated in time by approximately $\ln(J_1/J_d)/\mu$, independent of $n$.
The minima in \subref{fig:single-distance}{(b)} are indeed evenly spaced, consistent
with this picture.

\section{Error Minimization}
\label{sec:error-minimization}
In this appendix we describe approaches to minimize error within our
simulations. Several factors were considered to ensure that the observed effects
arose from the chaotic dynamics rather than numerical error. We describe the two
main numerical choices used to control these errors.

\subsection{Time Step Dependence}
\label{sec:time-step}
The use of RK4 requires choosing a time step small enough to control
discretization error while remaining computationally practical. A time step that
is too large would produce too much error causing inspection of chaos in the
system to be difficult. A time step too small introduces roundoff error. To
guide our choice of the correct time step we consider the expected value of
$D(0,0)$. The perturbed spin follows from the Rodrigues rotation formula,
\begin{equation}
    \vec{S}'_0 = \cos\epsilon\,\vec{S}_0
    + (1-\cos\epsilon)(\vhat{n}\cdot\vec{S}_0)\,\vhat{n}
    + \sin\epsilon\,\vhat{n}\times\vec{S}_0 .
    \label{eq:rodrigues}
\end{equation}
The last term is orthogonal to $\vec{S}_0$, and for a random axis
$\avg{(\vhat{n}\cdot\vec{S}_0)^2}=1/3$, so
\begin{equation}
    D(0,0) = \tfrac{2}{3}(1-\cos\epsilon) = \frac{\epsilon^2}{3} + O(\epsilon^4).
    \label{eq:initial-decorrelator}
\end{equation}
At early times, before exponential growth sets in, $D(0,t)$ should remain on
this plateau; time steps that are too large produce spurious growth instead.

\begin{figure}[t]
    \centering   
    \includegraphics[width=\columnwidth]{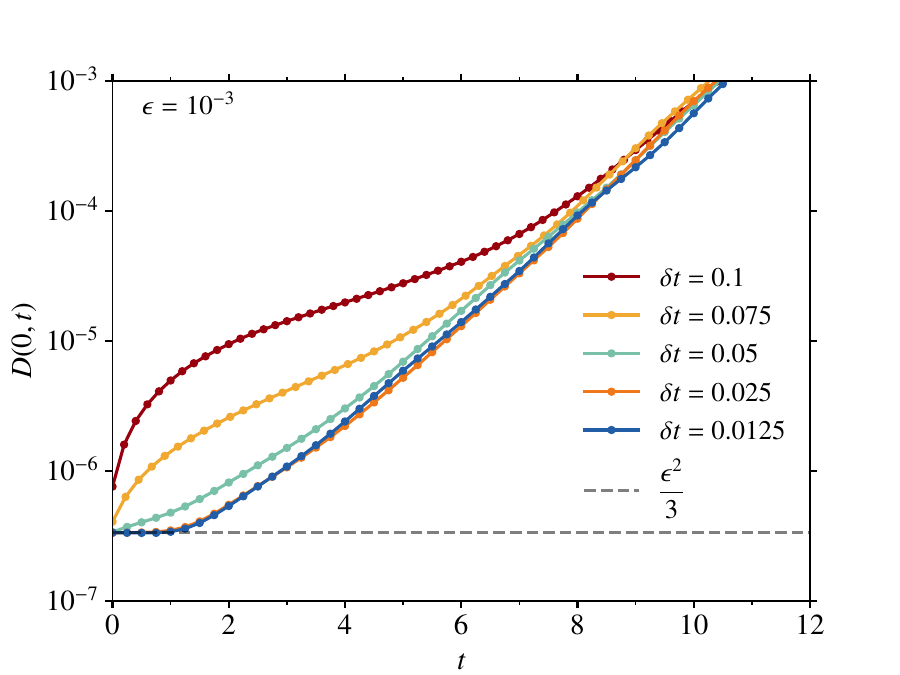}    
    \caption{Decorrelator at the perturbed site, $D(0,t)$, for $\epsilon =
    10^{-3}$ and several time steps $\delta t$. The dashed line marks the
    initial value $\epsilon^2/3$ [\cref{eq:initial-decorrelator}]. As $\delta t$
    decreases, the form of $D(0,t)$ becomes more stable. At $\delta t = 0.025$,
    we see consistent behavior that agrees with smaller time steps.}
    \label{fig:time-step}
\end{figure}

In \cref{fig:time-step} we test a range of time steps. At larger time steps,
$\delta t = 0.1$, $0.075$, and $0.05$, RK4 produces spurious early-time growth
of $D(0,t)$. At $\delta t = 0.025$ and $0.0125$, $D(0,t)$ remains at its initial
value until exponential growth sets in. Since $\delta t = 0.025$ gives the same
results as $\delta t = 0.0125$ at lower computational cost, we use it as the
fixed time step for our simulations. Although \cref{fig:time-step} shows only
$D(0,t)$, we have also compared full long-range simulations at $\J=10^{-3}$ with
$\delta t=0.01$ and $0.025$ and found no appreciable change in the arrival-time
fronts extracted at $\chi=10^{-3}$. Since the physical initial decorrelator
scales as $D(0,0)\sim\epsilon^2$ while the RK4 discretization produces an
$\epsilon$-independent numerical error floor, resolving the plateau for smaller
perturbations requires a correspondingly smaller time step.

\subsection{Choice of Integrator}
\label{sec:integrator}
\begin{figure}[t!]
    \centering
    \includegraphics[width=\columnwidth]{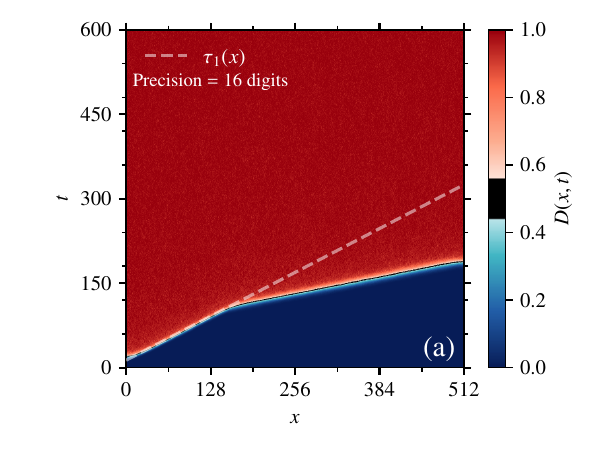}
    \includegraphics[width=\columnwidth]{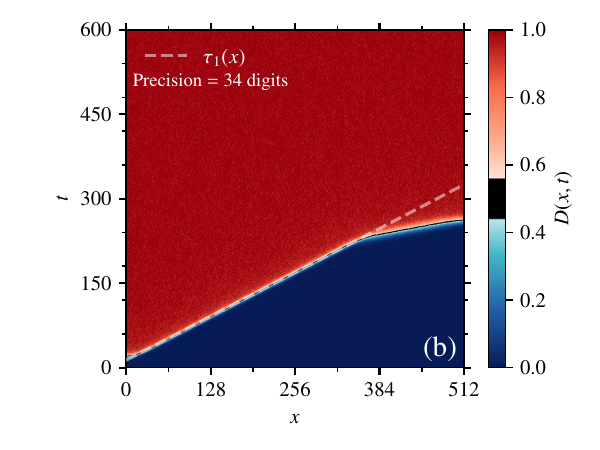}
    \centering
    \includegraphics[width=\columnwidth]{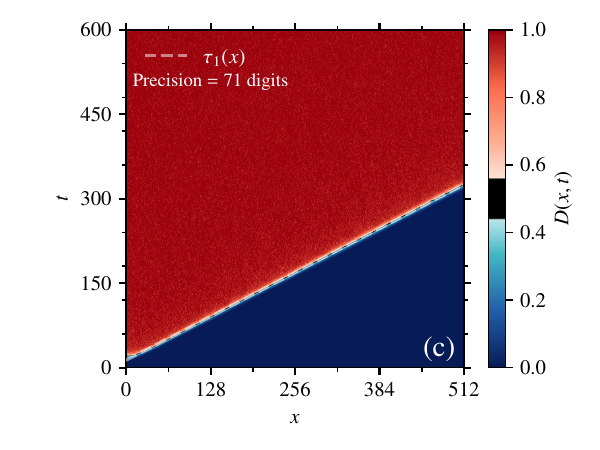}
    \caption{Simulated data $D(x,t)$ using the fourth-order Suzuki--Trotter
    integrator for a perturbation of size $\epsilon = 10^{-3}$ in a chain of
    length $N = 1024$ with nearest-neighbor interactions. Different precisions
    were used to reproduce the nearest-neighbor cone found using RK4. For this
    implementation, 71-digit (octuple) precision reproduces the nearest-neighbor
    cone, whereas the lower precisions shown do not. RK4 reproduces the same
    front in double precision.}
    \label{fig:symplectic-precision}
\end{figure}

We also tested a fourth-order Suzuki--Trotter integrator based on alternating
exact rotations of the two sublattices of the nearest-neighbor chain, using the
fourth-order composition described in Ref.~\cite{yoshida1990symplectic}. For
this implementation, substantially higher arithmetic precision was needed to
reproduce the nearest-neighbor front than for RK4
(\cref{fig:symplectic-precision}). These results made this integrator unsuitable
for the long-range simulations.

\section{Nearest-Neighbor Analysis}
\label{sec:nn-analysis}

\begin{figure}[b!]
    \centering   
    \includegraphics[width=\columnwidth]{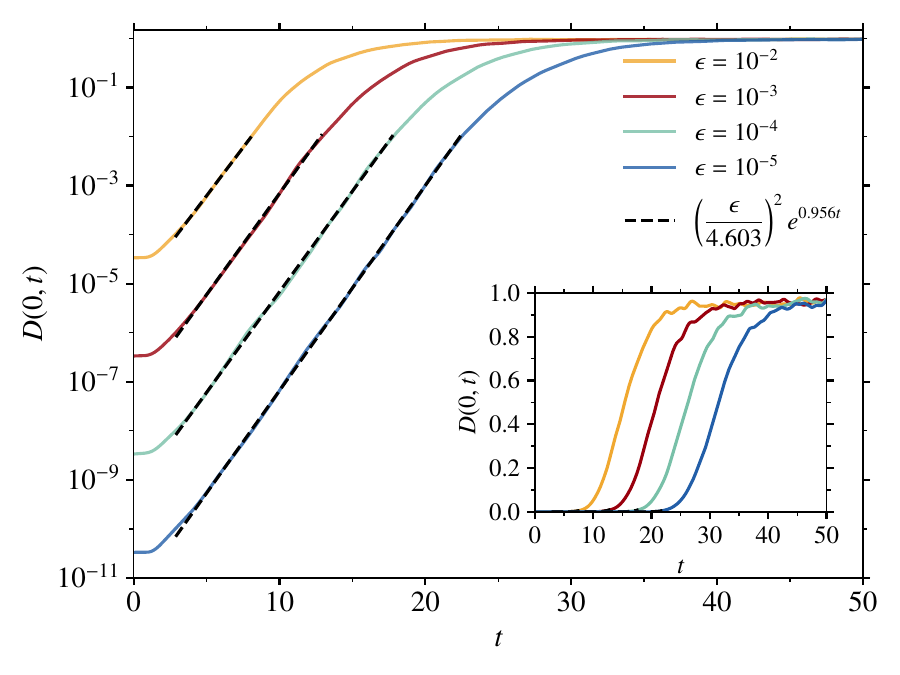}
    \caption{Decorrelator at the perturbed site, $D(0,t)$, for
    $\epsilon=10^{-2}$ to $10^{-5}$. Dashed lines show fits of the form
    $A\epsilon^2e^{2\mu t}$ [\cref{eq:nn-growth-form} at $x=0$], which describe
    the data for $\epsilon^2\lesssim D\lesssim10^{-2}$. Inset: the same data on
    a linear scale, showing saturation at $D\approx1$.}
    \label{fig:nn-epsilon}
\end{figure}

In this Appendix, we reproduce the results of Ref.~\cite{das2018light} to ensure
that they can be used as a foundation for the long-range case. In
\cref{fig:nn-epsilon}, we show the decorrelator at the perturbed site, $D(0,t)$,
for several values of $\epsilon$ and compare it with \cref{eq:nn-growth-form}
evaluated at $x=0$.

First, as $\epsilon$ decreases, $D(0,t)$ shifts to later times, delaying the
onset of decorrelation. The perturbation size does not affect the rate of
propagation of chaos, which remains fixed at the butterfly velocity $v_b$ for
all $\epsilon$. Second, the decorrelator scales as $\epsilon^2$, so the
asymptotic front shifts in time by $\ln(1/\epsilon)/(2\mu)$.\footnote{Setting
$D_1(x,t)=\chi$ in \cref{eq:nn-growth-form} gives
$\tau_1(x)=\tau_0/2+(\tau_0^2/4+x^2/v_b^2)^{1/2}$, with $\tau_0$ as in
\cref{eq:seed-arrival-time}. At the perturbed site, $\tau_1(0)=\tau_0$ shifts by
$\ln(1/\epsilon)/\mu$, while at large $x$, $\tau_1(x)\simeq\tau_0/2+x/v_b$
shifts by half as much.} Matching the data to \cref{eq:nn-growth-form} fixes the
normalization at $A\approx1/21.2$ and the growth rate at $\mu = \muval$. This
value of $\mu$ differs from the $\approx 0.494$ reported in \citet{das2018light}
by a few percent.

The growth form of Ref.~\cite{das2018light} is valid only at early times, while
$D\ll1$, so the front must be extracted at a threshold where the perturbed and
unperturbed chains have only begun to diverge. The data and model agree for
$\epsilon^2\lesssim D\lesssim10^{-2}$. Increasing the threshold to be of order
unity leads to a fit that does not align with the majority of the data. 

Accordingly, throughout we extract the front at $\chi = 10^{-3}$, which lies
within the range where this model holds.

\bibliography{main}
\end{document}